\documentclass[journal,twoside,web]{ieeecolor}
\usepackage{tmi}
\usepackage{cite}
\usepackage{amsmath,amssymb,amsfonts}
\usepackage{algorithmic}
\usepackage{graphicx}
\usepackage{textcomp}

\usepackage{array}
\usepackage{bm}
\usepackage{bbm}
\usepackage{multirow}

\def\BibTeX{{\rm B\kern-.05em{\sc i\kern-.025em b}\kern-.08em
    T\kern-.1667em\lower.7ex\hbox{E}\kern-.125emX}}
\begin{document}
\title{Interactive Radiotherapy Target Delineation \\with 3D-Fused Context Propagation}
\author{Chun-Hung Chao, Hsien-Tzu Cheng, Tsung-Ying Ho, Le Lu, and Min Sun

\thanks{C.-H. Chao, H.-T. Cheng, and M. Sun is with the Department of Electrical Engineering, National Tsing Hua University, Hsinchu, Taiwan ROC (e-mail: raul.c.chao@gmail.com; hsientzucheng@gapp.nthu.edu.tw; sunmin@ee.nthu.edu.tw).}
\thanks{T.-Y. Ho is with the Department of Nuclear Medicine, Chang Gung Memorial Hospital, Taoyuan, Taiwan ROC (e-mail: albertyho@gmail.com).}
\thanks{L. Lu is with the PAII Inc Bethesda Research Lab, Bethesda, MD 20817 USA (e-mail: tiger.lelu@gmail.com).}}

\maketitle

\begin{abstract}
Gross tumor volume (GTV) delineation on tomography medical imaging is crucial for radiotherapy planning and cancer diagnosis. Convolutional neural networks (CNNs) has been predominated on automatic 3D medical segmentation tasks, including contouring the radiotherapy target given 3D CT volume. While CNNs may provide feasible outcome, in clinical scenario, double-check and prediction refinement by experts is still necessary because of CNNs' inconsistent performance on unexpected patient cases. To provide experts an efficient way to modify the CNN predictions without retrain the model, we propose 3D-fused context propagation, which propagates any edited slice to the whole 3D volume. By considering the high-level feature maps, the radiation oncologists would only required to edit few slices to guide the correction and refine the whole prediction volume. Specifically, we leverage the backpropagation for activation technique to convey the user editing information backwardly to the latent space and generate new prediction based on the updated and original feature. During the interaction, our proposed approach reuses the extant extracted features and does not alter the existing 3D CNN model architectures, avoiding the perturbation on other predictions. The proposed method is evaluated on two published radiotherapy target contouring datasets of nasopharyngeal and esophageal cancer. The experimental results demonstrate that our proposed method is able to further effectively improve the existing segmentation prediction from different model architectures given oncologists' interactive inputs.
\end{abstract}

\begin{IEEEkeywords}
Radiotherapy planning, interactive segmentation, esophageal cancer, nasopharyngeal cancer, convolutional neural network
\end{IEEEkeywords}

\section{Introduction}
\label{sec:introduction}

\subsection{Motivation}
\IEEEPARstart{G}{ross} Tumor Volume (GTV) delineation on the radiotherapy planning computerized tomography (CT) scan image is a crucial process for tumor treatment. For most radiation oncologists, contouring 3D GTV as precisely as possible is part of their daily practice. The challenge of this task is the non-clarity of where tumor–normal tissue boundaries occur, especially for heterogeneous and infiltrative tumors. Automatic segmentation for the radiotherapy target \cite{el2007concurrent,pasquier2007automatic,ciernik20073d,zaidi2010pet} has been proposed before the thriving of deep neural networks.



In the clinical scenario, when given an initial radiotherapy target volume, the oncologists usually need to refine and edit it slice-by-slice in axial-view. As the anatomical structure gradually changes along the z-axis, the oncologists must delineate the corresponding target contours, which are also gradually changing, throughout all the slices. Such process is tedious and requires persisting focus; to form a desired target volume, they must account for subtle differences in neighboring slices even though the target contours on each slice could have almost the exact same appearance as those adjacent.
Motivated by the oncologists' practice, we propose to have an interactive method to propagate the revised delineation results in 3D manners, mitigating their burden to slice-by-slice investigate into the whole CT volume.

Automatic medical segmentation by deep neural networks (DNN) has been predominated and proven to have better performance than the traditional machine learning methods. Because of the complexity and diversity of the real world medical cases, the automatic medical systems often serve as assistant to medical experts. Therefore, during segmentation process in intricate medical diagnosis such as treatment target contouring for radiotherapy, experts may constantly revise the result predicted by automatic segmentation systems. As more DNN segmentation models emerged to assist diagnostic tasks, the demand of flexible and efficient tools to interact with results predicted by DNN increases simultaneously.

Similar to numerous of medical segmentation applications, radiotherapy target delineation is a 3D segmentation task. However, most of the proposed interactive approaches \cite{Sun2018InteractiveMI,amrehn2017ui,Wang2016SlicSegAM,zhu2014effective,Top2011ActiveLF,acuna2018efficient,hong2020structure,ling2019fast} are based on 2D segmentation, which often require specific user-manual control points, extreme points, or scribbles. In radiotherapy target contouring, the cost of directly drawing the perimeter of target regions in a 2D axial-view slice is arguably affordable or even lower than user-manual foreground scribbles. In fact, the cross-slices annotation process, i.e., contouring 3D radiotherapy target considering not only the axial view but sagittal view and coronal view, could be a more significant burden for the radiation oncologists. To tackle this, our method is able to fuse and propagate a re-annotated slice to its 3D context entirely, mitigating the cost of manual investigation into other slices.

As different segmentation models could be feasible for different clinical scenarios, the interactive method should better be decoupled with the DNN model which provides baseline segmentation. Recent DNN interactive segmentation methods \cite{curveGCN,bredell2018iterCNN} require customized built-in architecture to handle the feed-back loop for the revised model output. Instead, our 3D interactive method is able to serve as a plug-in module appending after any 3D segmentation DNN, updating its model output accordingly while additional training or modification on original segmentation DNN is not required. By given edited slices at any relative position in a 3D image volume, the proposed method incorporates the spatial relationship between the edited slices and the other remain slices by propagating to its 3D context and fusing the updated 3D feature maps to produce refined segmentation. Our experiments on nnUNet~\cite{isensee2019nnUNet} and PSNN~\cite{jin2019PSNN} demonstrate that our interactive method is able to work on different 3D segmentation DNNs on different radiotherapy tasks.

Inspired by previous works exploited backpropagation for activations for the purpose of visualization, interacting with neural networks, depth reconstruction, etc., we adopt similar backpropagation scheme for interactive segmentation. In contrast to typical backpropagation acting on parameters, which is used for training network, backpropagation for activations is a process that can be used for transporting data backwardly through network layers. In our work, interaction information form the user edited slice is conveyed back to the latent space with such backpropagation technique while the feature map is the optimization target. For the optimized slices in the 3D feature map volume, the contours on their neighboring slices may also be improved due to the overlapped receptive fields and similar geometric appearance. However, for the farther slices, the contours on them won't be modified or would be corrupted since the receptive fields are not overlapped or the backpropagation process makes the whole feature map out of the distribution. Thus for the farther slices, we propose to use a fusion network which takes the feature maps before and after updated as input to tackle the drawbacks of the backpropagation process and regularize the updated predictions.

In this paper, we propose a novel interactive 3D radiotherapy target delineation framework that can work with main stream 3D deep learning segmentation networks. Moreover, our framework addresses the limitations of the direct backpropagation for activations update, which makes more use of each user interaction. We conducted our experiments on two GTV segmentation dataset, from the experimental results, $19.6\%$ and $9.7\%$ of whole GTV volume performance boosts can be observed with single slice edited. In addition, we can reach an average of 0.8 dice score on the testing set with merely 4 and 1 slices editing respectively. With our framework, we can increase the flexibilty of deep learning method at the clinical scenario for radiotherapy target delineation, and further reduce the human power and time required to be invested in such labor-intensive task.



\begin{figure}[t!]
\begin{center}
\includegraphics[width=1.0\columnwidth]{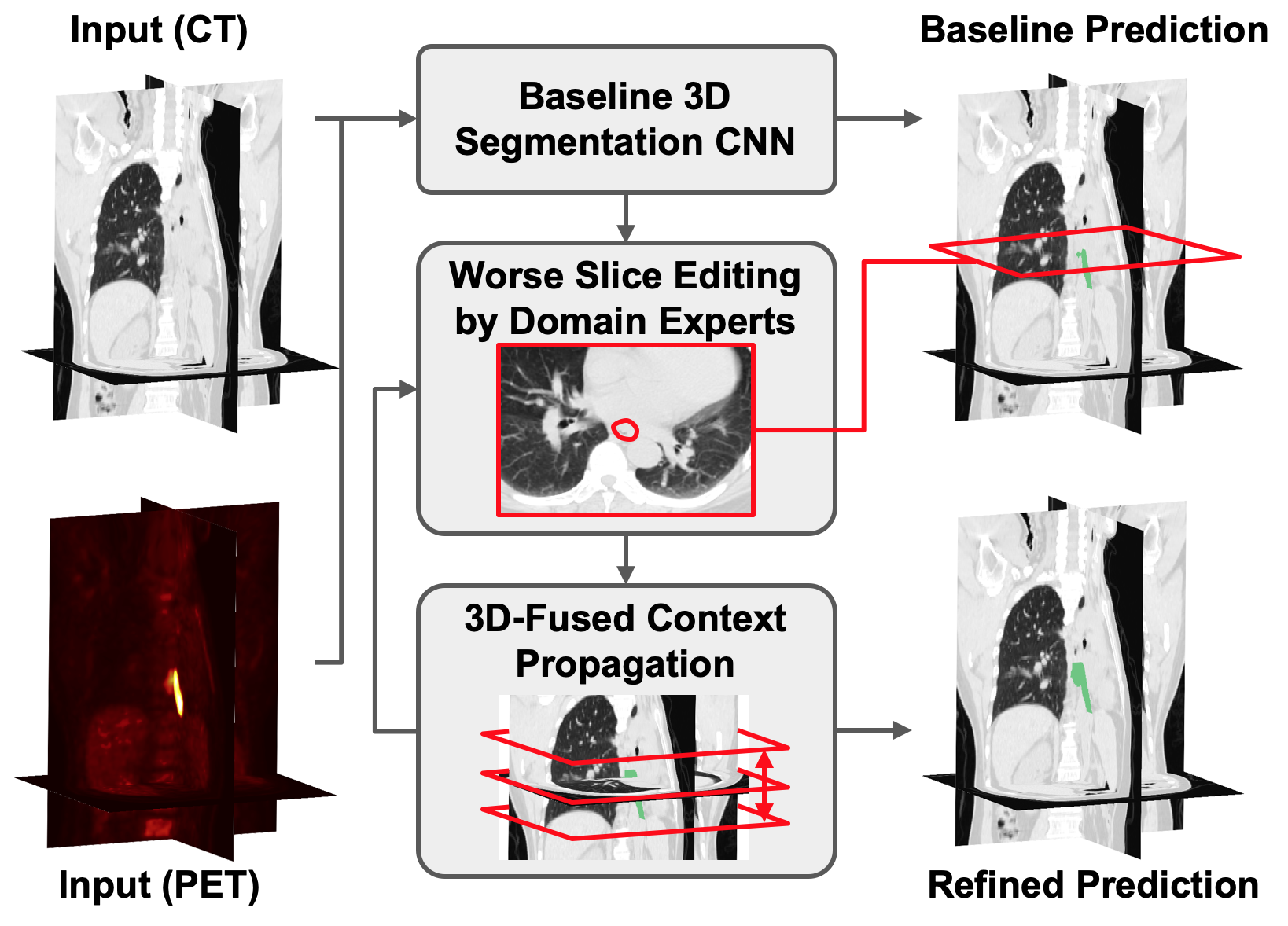}
\end{center}
\caption{Interaction pipeline overview. Given a CT or paired CT-PET image volume, a off-the-shelf 3D segmentation CNN would first generate rudimentary radiotherapy target contour predictions. Then, the radiation oncologist reviews the prediction and perform editing on the slice with worst performance in the volume and our interactive framework takes the corrections as its input and improve the whole volume from the modification on the single slice. This interactive procedure can be performed iteratively until the oncologists consider the produced target contour delineations eligible for the following radiation treatment process.}
\label{fig.concept}
\end{figure}

\subsection{Related Work}

\subsubsection{Backpropagation for Activations}
Backpropagation is usually used for training neural network by computing the gradient with respect to network parameters and update the parameters accordingly. Aside from the usage of optimizing network parameters, backpropagation can also be used to update feature maps or input image under the condition of fixed network parameters. Simonyan et. al \cite{simonyan2013deep} propose to use vanilla backpropagation as an gradient-based approach to inverse the data flow and visualize the concepts of a class learned by CNN by updating the input image to maximize the classification score. Rupprecht et. al \cite{rupprecht2018guide} use backpropagation to estimate the optimal guiding parameters from given hints for feature map manipulation, enabling users to modify network's activations and interact with CNN. Also, Wang et. al \cite{wang2019plug} propose to use backpropagation to update feature map to propagate sparse input depth data to further improve depth prediction from original model. Similar to \cite{rupprecht2018guide} and \cite{wang2019plug}, in \cite{jang2019interactive}, a backpropagation refinement scheme is proposed to refine segmentation mask with extra user clicks provided. However, none of the mentioned works is applied to the 3D image volume, which is common in medical imaging. In our work, we explore the usability of the backpropagation for activations for 3D medical image, and address the limitations of not able to improve the performance on farther slices.

\subsubsection{3D Biomedical Image Segmentation}
In clinical practice, especially oncology, physicians heavily rely on 3D medical images to make diagnoses or treatment plans. Various CNN based image segmentation methods have been de,veloped to give direct or indirect assistance in medical procedures and related biological researches. 2D based segmentation methods \cite{song2015lung,makropoulos2014automatic,tsai2003shape,cherukuri2017learning,huang2014brain,chen2017deeplab,li2018h,gu2019net} has first been introduced to process volumetric data in slice-wise manners. 3D CNN such as 3D U-Net \cite{3DU-Net} and V-net\cite{VNet,gibson2018automatic} were developed to deal with volumetric contexts directly. \cite{Ren2018in} used 3D convolutional kernels for volumetric segmentation of structures and tissues in the head and neck. A 3D-based CNN is also used in the coarse-to-fine framework \cite{zhu20173d} to segment the pancreas from CT scans. Recently, Yousefi et al.~\cite{yousefi2018esophageal} proposed to use a 3D U-shape CNN to predict GTV. Jin et al.~\cite{jin2019accurate} use 3D network fusing CT and PET modalities to predict esophageal GTV. Our method focuses on further improving the prediction from these methods, to have the performance fulfill the clinical needs with user interactions, which increase the usability and flexibility of the 3D medical image segmentation frameworks, especially those for radiotherapy targets.

\subsubsection{Interactive Segmentation}
To generate an accurate treatment plan which meets radiologists' knowledge and experience, cooperation with experts is one of the most crucial issues in our application. In fact, user-assisted segmentation for the 3D medical image has been studied for years \cite{criminisi2008geos, Top2011ActiveLF, zhu2014effective, Wang2016SlicSegAM}. Recently, factoring user input into the deep learning model has become a popular scheme. DeepIGeoS \cite{Wang2018DeepIGeoSAD} adds user interaction to an additional channel of CNN via geodesic transform. UI-Net \cite{amrehn2017ui} includes a fully convolutional neural network (FCN) which can iteratively accept user's hints represented by seed points. Image-specific fine-tuning and a weighted loss function \cite{wang2018interactive} are also used to make CNN model adaptive to a specific test image with user inputs. Iterative interaction training (interCNN) \cite{bredell2018iterCNN} is proposed to retrain segmentation based on users' scribble on the network prediction. While the previous methods require the user to edit each predicted image, our model aims to propagate information back into high dimensional feature space and then refine the prediction based on it. Furthermore, unlike previous interactive methods that require specific network structures, our method is able to plug into any 3D CNN segmentation models without retraining and overwriting the parameters of baseline CNN.

\subsection{Contributions}

Our contributions:
\begin{enumerate}
    \item We propose a backpropagation-based interactive network as a generic method to be applied on 3D medical segmentation models for simple yet efficient human editing while the original 3D medical segmentation models would be not be retrained or modified.
    \item We discuss the usage of backpropagation for activations for 3D medical image and tackle the problem that direct backpropagation method has limited improved range along z axis.
    \item For medical target without clear foreground boundary, such as GTV, our method is able to handle medical implicated contours better than previous interactive methods since the information encoded in the high dimension feature space by the prediction model is reused.
    \item Our method further improve the state of the art models performance on two GTV delineation tasks while only a slice is edited by the user: nasopharynx cancer with 19.63\%, and esophagus cancer with 9.74\%.
\end{enumerate}

\section{Method}
\label{sec:method}

\begin{figure*}[!t]
\begin{center}
\includegraphics[width=0.85\linewidth]{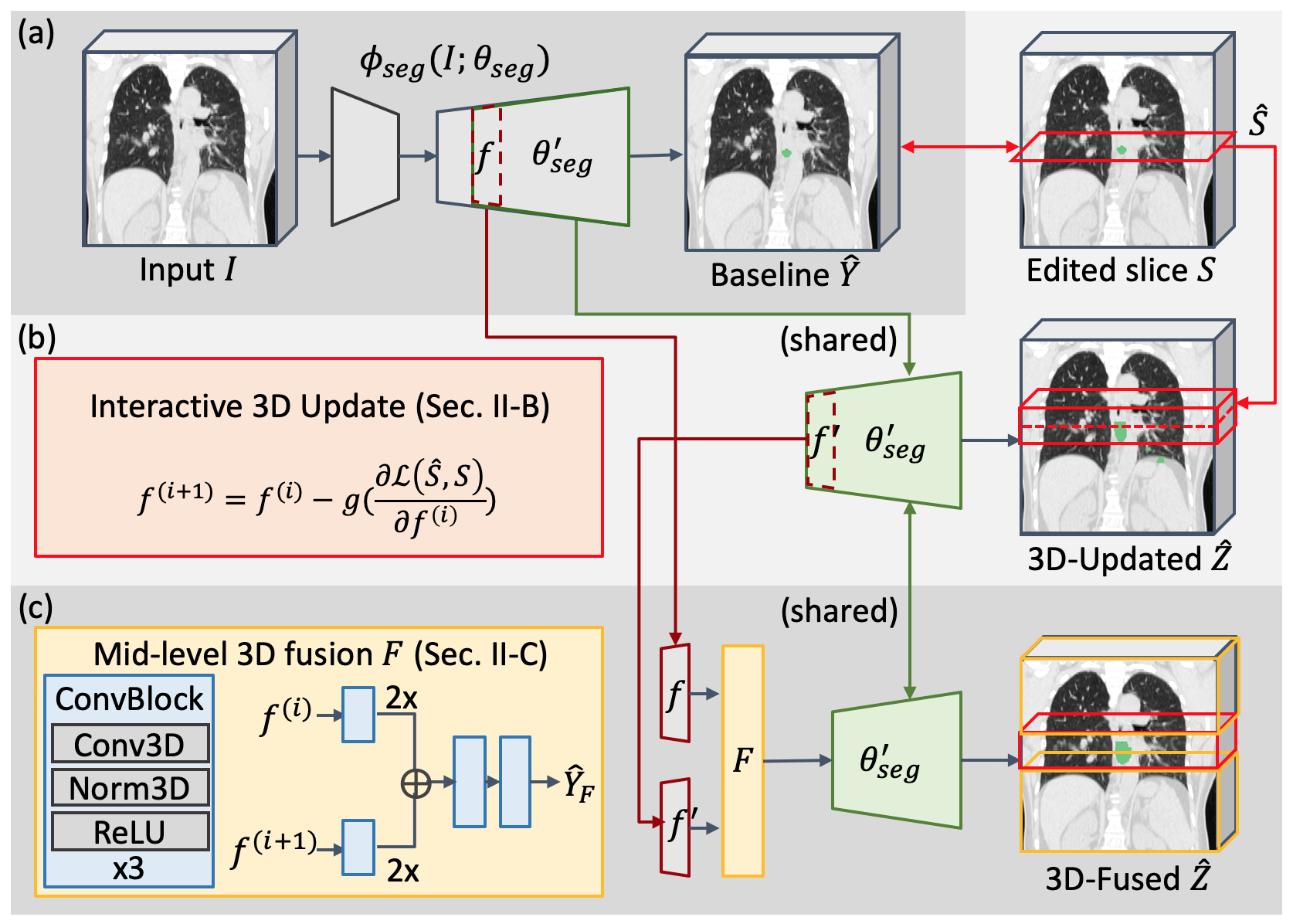}
\caption{The proposed method. (a) Baseline 3D segmentation model. Given a segmentation model with input volume $I$, the initial prediction $\hat{Y}$ is generated via $\phi_{seg}$ with model weight $\theta_{seg}$. (b) Interactive 3D update by back-propagation of activation. With edited slice $S$ modified from the initial prediction, the response back-propagates through decoder $\theta^{'}_{seg}$, updating $f^{'}$ to inference 3D-Updated $\hat{Z}$. (c) Mid-level 3D fusion by baseline and updated feature maps. The feature from the fusion module $F$, fused by the original feature $f$ and the updated feature $f^{'}$, inference through the decoder $\theta^{'}_{seg}$ again for the final 3D-Fused $\hat{Z}$. Note that in all stages (a, b, and c), the decoder model weight $\theta^{'}_{seg}$ has remained unchanged.}
\label{fig:method}
\end{center}
\end{figure*}

\begin{figure}[t!]
\begin{center}
\includegraphics[width=1.0\columnwidth]{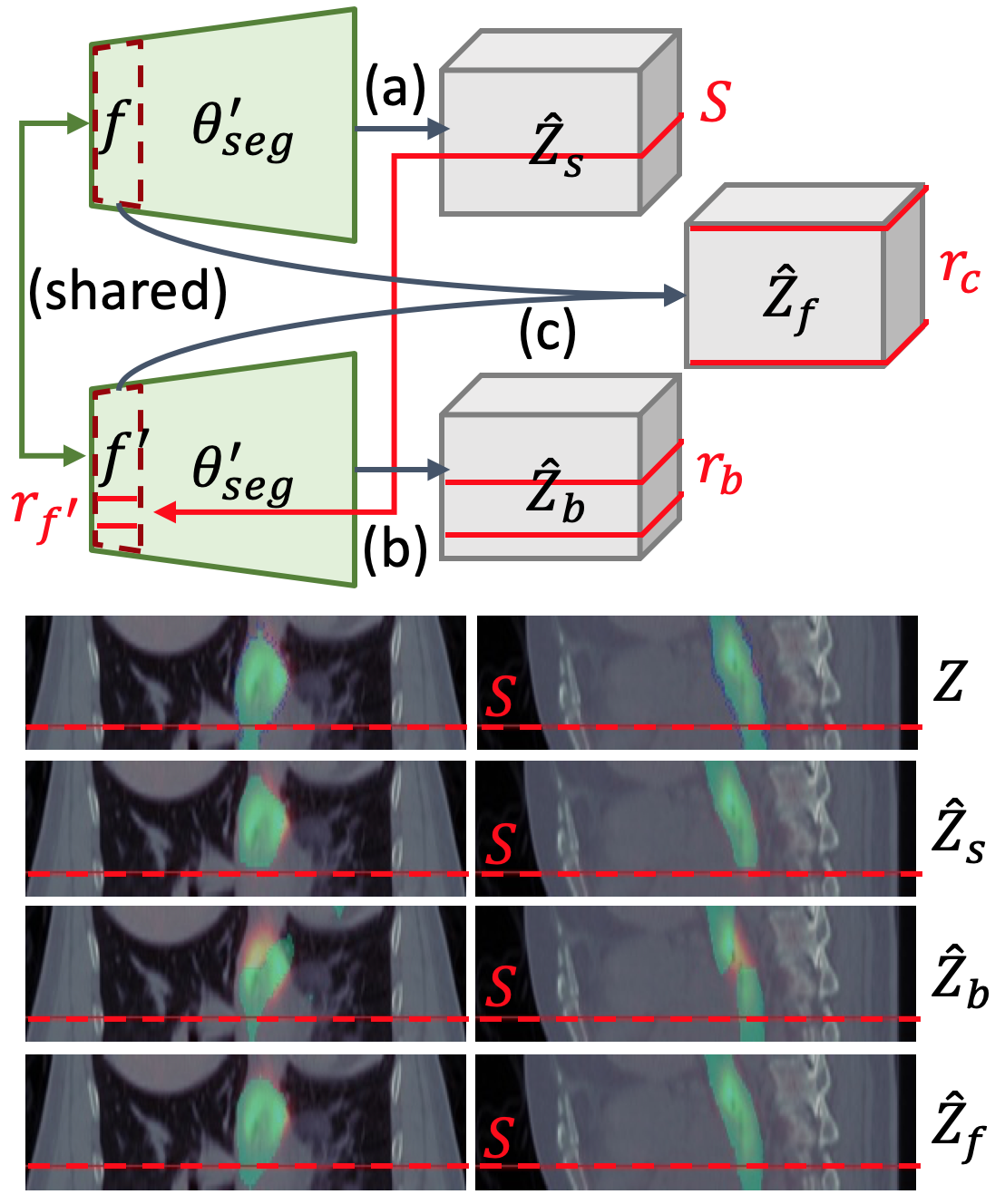}
\end{center}
\caption{The back-propagation by activation might lead to prediction corruption because of its limited receptive field (RF). (a), (b), and (c) refer to Fig.~\ref{fig:method}. After inference via (a) producing $\hat{Z_s}$, BP from $S$ affects $f'$ with RF $r_{f'}$. Interactive updated $f'$ then inference via (b) for $\hat{Z_b}$ with affected RF $r_{b}$. The mid-level fusion (c) will cover RF $r_{c}$ when producing $\hat{Z_f}$}
\label{fig.bp_corruption}
\end{figure}

\subsection{Preliminaries and Notations}

We define a RTCT or registered RTCT/PET pairs image volume as $\bm{I} \in \mathbb{R}^{C \times D \times H \times W}$, where $D$, $H$, and $W$ is the depth (distance in the Z direction), height, and width of the image, and $C \in \{1, 2\}$ is the number of channel depends on the image modality, i.e., RTCT or RTCT/PET. Then we can feed the image $\bm{I}$ into a segmentation model $\phi_{seg}$ to obtain the initial prediction volume $\hat{\bm{Y}} \in \mathbb{R}^{D \times H \times W}$ and a feature map volume $\bm{f} \in \mathbb{R}^{C' \times D' \times H' \times W'}$ at a specific layer. User can edit a slice in $\hat{\bm{Y}}$, and we are able to update feature map $\bm{f}$ to $\bm{f}^{'}$ according to the edited slice via backpropagation. Then our fusion model $\phi_{fusion}$ takes both $\bm{f}$ and $\bm{f}^{'}$ as input to produce the modified prediction $\hat{\bm{Z}} \in \mathbb{R}^{D \times H \times W}$ based on user interaction. Eventually, we evaluate the performance according to $\hat{\bm{Z}}$ and label $\bm{Y} \in \mathbb{R}^{D \times H \times W}$.

\subsection{Interactive Update Module}\label{sec:backprop}
Like typical CNN, we obtain the segmentation prediction with our segmentation model $\phi_{seg}$ through a feed-forward inference path
\begin{equation}\label{eqn:seginf}
    \hat{\bm{Y}} = \phi_{seg}(\bm{I};\theta_{seg}),
\end{equation}
where the segmentation model $\phi_{seg}$ is parameterized by $\theta_{seg}$ and the input image $\bm{I}$. At this moment, user might feel unsatisfactory about a slice $\bm{s}$ in $\hat{\bm{Y}}$ and would like to make modification to it. From Eq.\ref{eqn:seginf}, we have two choices to incorporate and force the prediction to align with the user's interaction: either by retraining $\phi_{seg}$ to obtain new $\theta_{seg}$ or updating $\bm{I}$. However, re-training the model may cause corruptions on the previous learned knowledge during training and it is also impractical to train a new model every time user edit a slice. Since our segmentation model is a feed-forward network, a mechanism that can invert the data flow and transporting the user interaction information backwardly is required for input image updating. To achieve this goal, we can leverage the backpropagation algorithm, which is originally to compute the gradient of the loss function with respect to the parameters from the last layer to the first by the chain rule. Considering the computational costs of backpropagation, we cut down the backpropagation path and update the intermediate feature map instead of the input image. Under this setting, we first calculate the error between the user edited slice and the corresponding slice in the prediction produced from the feature map. Then, we update the feature map according to the gradient of the error with respect to the feature map, which is obtained via the backpropagation algorithm. This process is conducted repeatedly until a certain stopping criteria, such as a threshold for error or a number of iterations, is reached.
\begin{align}
    \hat{\bm{Y}} &= \phi_{seg}^{'}(\bm{f};\theta_{seg}^{'})\\
    \bm{f}^{(i+1)} &= \bm{f}^{(i)} - g(\dfrac{\partial \mathcal{L}(\hat{\bm{S}}, \bm{S})}{\partial \bm{f}^{(i)}}) \label{eqn:backprop}
\end{align}
$\phi_{seg}^{'}$ and $\theta_{seg}^{'}$ denotes the rest part of the model and its parameters where $\bm{f}$ is directed to in $\phi_{seg}$. Eq.\ref{eqn:backprop} shows how the feature map $\bm{f}$ at iteration $i$ is updated. $\bm{S}$ and $\hat{\bm{S}}$ are the user edited slice and the corresponding slice in $\hat{\bm{Y}}$ respectively. $\mathcal{L}$ is the loss function for segmentation performance. $f^{i}$ is updated based on the optimization algorithm $g$. Note that $g$ can also be second or higher order optimization algorithm as long as it can get its required term from the first order derivatives.

In addition to enforcing the prediction to match the user interaction on the modified slice, the backpropagation scheme is also capable of altering the adjacent slices. Generally, nearby image slices have similar anatomical structure and thus the radiotherapy contours on them would also be similar. Likewise, if the model fails to make good prediction for a slice, it also fails for the neighboring slices. As shown in Fig.\ref{xxx}, the backpropagation conveys the user editing information to the corresponding section in the feature map, which also falls inside the receptive field of the neighboring slices. As the features in the receptive field changed in this way, the activations are also influenced and the new predictions would become similar to the user edited slice. According to our experiment, we can substantially improve the prediction quality on the neighboring slices.

\subsection{Mid-level Fusion Module}

As described in Sec.\label{sec:backprop}, the backpropagation for activations technique makes great improvement to the neighboring slices of the selected slices. However, for the farther slices, the updated features might not fall in or be a little part of their receptive fields. What's worse, for some network architectures and their training processes which are relatively sensitive, externally manipulated features would be considered as the out-of-distribution data for the network layers and consequently causes the following activations and the final contour predictions on the father slices be tragically corrupted (See Fig.\label{xxx}). Thus in addition to the direct backpropagation update for the neighboring slices, we utilize a fusion network for learn an adaptive fusion between the original feature map and the updated feature map for the non-neighboring ones. In this case, we expect the fusion module can
\begin{enumerate}
    \item propagate the user interaction information embedded in the feature map to the receptive field of farther slices
    \item properly integrate the original and updated feature map to further improve the predictions on the farther slices which are not edited.
\end{enumerate}

As presented in Fig.\label{xxx}, the original and updated feature map, $\bm{f}$ and $\bm{f}'$ are first fed into the two stream input convolution block. In the input convolution block, the strides of the first convolution kernels are set to $2 \times 2 \times 2$, which reduces the size of the input feature maps by $2 \times 2 \times 2$ but also increases the channel size by $2 \times 2 \times 2$. As the feature map dimension along z-axis is scaled down by 2, the convolution kernel operates in feature space is actually operating on more slices in image space when performing each convolution. Hence, the information can be propagated to farther slices in image space in this way. Via this process, the receptive field for single kernel is increased and higher information is generated. After the input convolution block extracted essential information from both $\bm{f}$ and $\bm{f}'$, the produced feature map are upsampled with trilinear interpolation and concatenated and directed to the join convolution block and the output convolution block. The first convolution kernels of join convolution and the output convolution block are designed to combine the cross-channel information and reduce the channel numbers. Then the output convolution block will yield the fused feature map which has identical size to the original feature map. Note that all the other convolution kernels not specifically mentioned above are all $3 \times 3 \times 3$ 3D convolution kernel and all the blocks in the fusion module are consisted of three Conv3D-InstanceNorm3D-ReLU sub-blocks.

Once we obtained the fused feature map, a clone of the rest part of the base segmentation model, $\phi_{seg}^{'}$, will take it as input. During the training, the loss is only computed on the non-neighboring slices. That is, the fusion module is dedicated for farther slices' contours prediction. The final modified prediction $\hat{\bm{Z}}$ can be illustrated as:
\begin{equation}
    \hat{\bm{Z}}(s) =
    \begin{cases}
    \phi_{seg}^{'}(\bm{f}^{'})(s), & \text{if s is a neighboring slice}\\
    \phi_{seg}^{'}(\phi_{fusion}(\bm{f}, \bm{f}^{'}))(s), & \text{if s is a farther slice}
    \end{cases}
\end{equation}
where $(\cdot)$ is the operation selecting a slice from an image volume with a given slice location $s$.

\subsection{Implementation Details}

Our method is evaluated on two GTV segmentation datasets, the Nasopharyngeal and the Esophageal Cancer GTV segmentation dataset, thus we respectively select one of the state-of-the-art methods for each dataset, the nnUnet and the PSNN, to be our $\phi_{seg}$. As $\phi_{fusion}$ share the same architecture with $\phi_{seg}$ in the succeeding pathway, $\phi_{seg}^{'}$, we adopt the same training loss used by $\phi_{seg}$ for corresponding $\phi_{fusion}$. For Nasopharyngeal dataset, we use a hybrid loss which is the combination of dice and cross-entropy loss while only dice loss is used for Esophageal dataset:
\begin{equation}
    \mathcal{L}_{dice} = -\dfrac{2\sum_{i \in Y^{'}}p_{i}y_{i}}{\sum_{i \in Y^{'}}p_{i} + \sum_{i \in Y^{'}}y_{i}},
\end{equation}
\begin{equation}
    \mathcal{L}_{CE} = -log(\dfrac{e^{z_c}}{\sum_{j \in \{0, 1\}}e^{z_j}}),
\end{equation}
\begin{equation}
    \mathcal{L}_{hybrid} = \mathcal{L}_{CE} + \mathcal{L}_{dice},
\end{equation}
where $p_i$ is softmax output for foreground prediction at pixel $i$ on farther slices $Y'$, $z_c$ is the output logits of target class $c \in \{0, 1\}$, which is either background or foreground.

Moreover, our system does support multi-slice editing with an iterative procedure. When slice $s$ is edited by the user, each of the neighbor slices of $s$ will be updates if it does not belong to the neighbor slices of any previous edited slice and the farther slices would be average with other farther slices prediction in preceding iterations.

For our feature map backpropagation updating part, we selected LBFGS\cite{liu1989limited} and Adam\cite{kingma2014adam} optimizer for Nasopharyngeal and Esaphageal dataset. The fusion module is optimized with Adam optimizer with learning rate set to $1 \times 10^{-3}$ and trained for 40 and 100 epochs on Nasopharyngeal and Esaphageal dataset on a single NVIDIA GTX 1080Ti GPU. The system is implemented with PyTorch \cite{paszke2019pytorch}.
\section{Experiments}
\label{sec:result}

\begin{table*}[ht!]
\caption{Nasopharyngeal GTV - Comparison of Methods.}
  \centering
    \small
    \begin{tabular}{|c|c|p{1.8cm}|p{1.8cm}|p{1.8cm}|p{1.8cm}|}
    \hline
    Volume-of-Interest & Method & DSC & 95\% HD & Sensitivity & Specificity\\
    \hline
    \hline
    \multirow{2}{*}{Worst Slice} & nnUNet Baseline & 49.80$\pm$0.14 & 9.712$\pm$2.578 & 0.578$\pm$0.130 & 0.996$\pm$0.002\\
    \cline{2-6}
    & Interactive 3D Update                    & \textbf{74.20$\pm$4.46} & \textbf{6.188$\pm$3.975} & \textbf{0.825$\pm$0.022} & \textbf{0.998$\pm$0.002}\\
    \hline
    \hline
    \multirow{3}{*}{GTV Volume} & nnUNet Baseline  & 61.29$\pm$1.34 & 7.611$\pm$1.838 & 0.665$\pm$0.099 & 0.996$\pm$0.001\\
    \cline{2-6}
    & Interactive 3D Update                    & 67.95$\pm$1.99 & 6.251$\pm$1.234 & 0.725$\pm$0.061 & 0.997$\pm$0.002\\
    \cline{2-6}
    & Mid-level 3D Fusion               & \textbf{73.32$\pm$1.55} & \textbf{5.575$\pm$0.721} & \textbf{0.784$\pm$0.012} & \textbf{0.997$\pm$0.000}\\
    \hline
    \end{tabular}%
  \label{tab.nasogtv-quant}%
\end{table*}%

\begin{table*}[ht!]
\caption{Esophageal GTV Result - Comparison of Methods.}
  \centering
    \small
    \begin{tabular}{|c|c|p{1.8cm}|p{1.8cm}|p{1.8cm}|p{1.8cm}|}
    \hline
    Volume-of-Interest & Method & DSC & 95\% HD & Sensitivity & Specificity  \\
    \hline
    \hline
    \multirow{2}{*}{Worst Slice} & PSNN Baseline & 40.98$\pm$2.91 & 6.625$\pm$1.925 & 0.482$\pm$0.025 & 0.999$\pm$0.000 \\
    \cline{2-6}
    & Interactive 3D Update                    & \textbf{86.64$\pm$1.12} & \textbf{2.076$\pm$0.559} & \textbf{0.886$\pm$0.013} & \textbf{0.999$\pm$0.000} \\
    \hline
    \hline
    \multirow{3}{*}{GTV Volume} & PSNN Baseline  & 75.45$\pm$0.93 & 8.457$\pm$1.560 & 0.710$\pm$0.012 & 0.999$\pm$0.000 \\
    \cline{2-6}
    & Interactive 3D Update                    & 73.55$\pm$1.32 & 25.494$\pm$2.947 & 0.755$\pm$0.027 & 0.998$\pm$0.000 \\
    \cline{2-6}
    & Mid-level  3D Fusion               & \textbf{82.80$\pm$0.19} & \textbf{5.531$\pm$0.938} & \textbf{0.805$\pm$0.008} & \textbf{0.999$\pm$0.000} \\
    \hline
    \end{tabular}%
  \label{tab.esogtv-quant}%
\end{table*}%

To demonstrate the effectiveness of our proposed interactive 3D propagation method, we conduct experiments on two published radiotherapy target delineation datasets. The experiments demonstrate the extent that the predictions of the state-of-the-art methods of two datasets can be further improved through our methods. We also simulate the clinical scenario and exhibit the average Dice similarity coefficient and qualitative results after multiple interactive modifications.

\subsection{Materials}
\subsubsection{GTV Segmentation for Nasopharyngeal Cancer}
This dataset is originally the training dataset of StructSeg 2019 challenge \cite{}. It is composed of 50 CT volumetric images from nasopharyngeal cancer patients and corresponding GTV segmentation masks provided by Zhejiang Cancer Hospital.
\subsubsection{GTV Segmentation for Esophageal Cancer}
This dataset is first published in \cite{jin2019accurate}. It contains CT-PET image pairs of 110 esophageal cancer patients diagnosed at stage II or later collected at Chang Gung Memorial Hospital. The CT and PET volumetric image scan of each patient is resampled to $1.0 \times 2.0 \times 2.5$ mm and registered with the B-spline based deformable registration algorithm \cite{rueckert1999nonrigid}. All the GTV target contours were delineated by two radiation oncologists during clinical workflow.

\subsection{Experimental Settings}
Similar to the user interaction simulation setting during training phase, we select the slice with largest Hausdorff distance to be the user edited slice. In our experiments, we evaluate our results produced from the single simulated user input within two different scope: the local subvolume and the whole GTV region to discuss the different impact of our method under single slice editing setting on different range, where the subvolume is composed of the edited slice and its neighboring 4 slices. We also conducted the ablation study of the multi-slice editing setting following the aggregation procedure described in Sec.\ref{} to demonstrate the expected number of extra interactions made by a radiotherapy oncologist to obtain a satisfactory GTV planning volume. In addition, we also show the variances in performance of backpropagation level from the last to the third-last layer of decoder. For all experiments, we performed 4-fold and 5-fold cross validation on the Nasopharyngeal and Esophageal dataset respectively.

\subsection{Evaluation Metrics}\label{subsec.eval_metrics} 
We refer to the metrics in \cite{taha2015metrics} and report our results with respect to the following four metrics, where $P$ and $Y$ denote the set of voxels of predicted and ground truth GTV contour respectively:

\noindent\textbf{Dice Similarity Coefficient.}
Dice Similarity Coefficient (DSC), also known by Srensen index or F1 score, is a statistic used for comparing the similarity of two samples and is the most used metric in validating medical volume segmentation. As shown in Eq. \ref{eq.dsc}, considering both prediction and target region, the intersection of them determines better DSC score:

\begin{equation}
    \text{Dice}(P, Y) = 2 \times \frac{|P \cap Y|}{|P| + |Y|}.
    \label{eq.dsc}
\end{equation}

\noindent\textbf{95\% Hausdorff Distance.}
The maximum Hausdorff distance is the maximum distance from the prediction voxel set to the label voxel set. 
\begin{equation}
    \text{HD}(P, Y) = \max(\max_{p \in P}\min_{y \in Y}d(p, y), \max_{y \in Y}\min_{p \in P}d(p, y)),
\end{equation}
where $d$ is the distance function.
The 95\% Hausdorff distance reports the 95th percentile of the distances to alleviate the impact of outliers.

\noindent\textbf{Sensitivity.}
Sensitivity (Eq. \ref{eq.sens}) measures that actual positives are not overlooked. This metric is the same as recall measurement, which shows the ratio of intersection over target region:

\begin{equation}
    \text{Sensitivity}(P, Y) = \frac{|P \cap Y|}{|Y|}.
    \label{eq.sens}
\end{equation}

\noindent\textbf{Specificity.}\\
Specificity (Eq. \ref{eq.spec}) assesses the statistics to which actual positives and negatives are classified. As precision measurement, the better ratio of intersection over prediction indicates better specificity:

\begin{equation}
    \text{Specificity} = \frac{|P^c \cap Y^c|}{|Y^c|}\text{ ,}
    \label{eq.spec}
\end{equation}
where superscript $c$ denotes the complement set.

\subsection{Quantitative Comparison}

Here we compare our method with baseline settings quantitatively with evaluation metrics described in Sec.~\ref{subsec.eval_metrics}. In addition to evaluating the whole GTV volume, we also evaluate the updating performance of nearby slices in worst slice volume-of-interest (VoI), i.e. the receptive field of $\phi^{'}_{seg}$. For result on Naso GTV data shown in Table.~\ref{tab.nasogtv-quant}, the interactive update performs better than the nnUNet baseline, demonstrating the effectiveness of propagating the edited slice the nearby slices. The mid-level 3D fusion $\phi_{fuse}$ perform even more better than $\phi^{'}_{seg}$ as updating prediction of not only the local slices but the far slices as a whole. On the other hand, the result of Eso GTV dataset is listed in Table.~\ref{tab.esogtv-quant}. In the worst slice VoI, the performance of $\phi^{'}_{seg}$ is significantly higher than the PSNN baseline. However, in the whole GTV point of view, the interactive 3D update causes slightly worse DSC and seriously worse 95$\%$ HD. This indicates the effect of prediction corruption mentioned in Fig.~\ref{fig.bp_corruption} due to the larger receptive field needed for the long Esophagus structure. By fusing the baseline and the interactively-updated feature maps, our $\phi_{fuse}$ outperforms the others by taking the local updated features and the larger GTV volume into account. The results in Table.~\ref{tab.nasogtv-quant} and Table.~\ref{tab.esogtv-quant} also demonstrate the generalizability of our proposed interactive model $\phi_{fuse}$ to be applied on data with anatomically different structures (nasopharyngeal and esophageal) to improve the performance from different baseline models.

\begin{figure*}[ht!]
\includegraphics[width=0.99\linewidth]{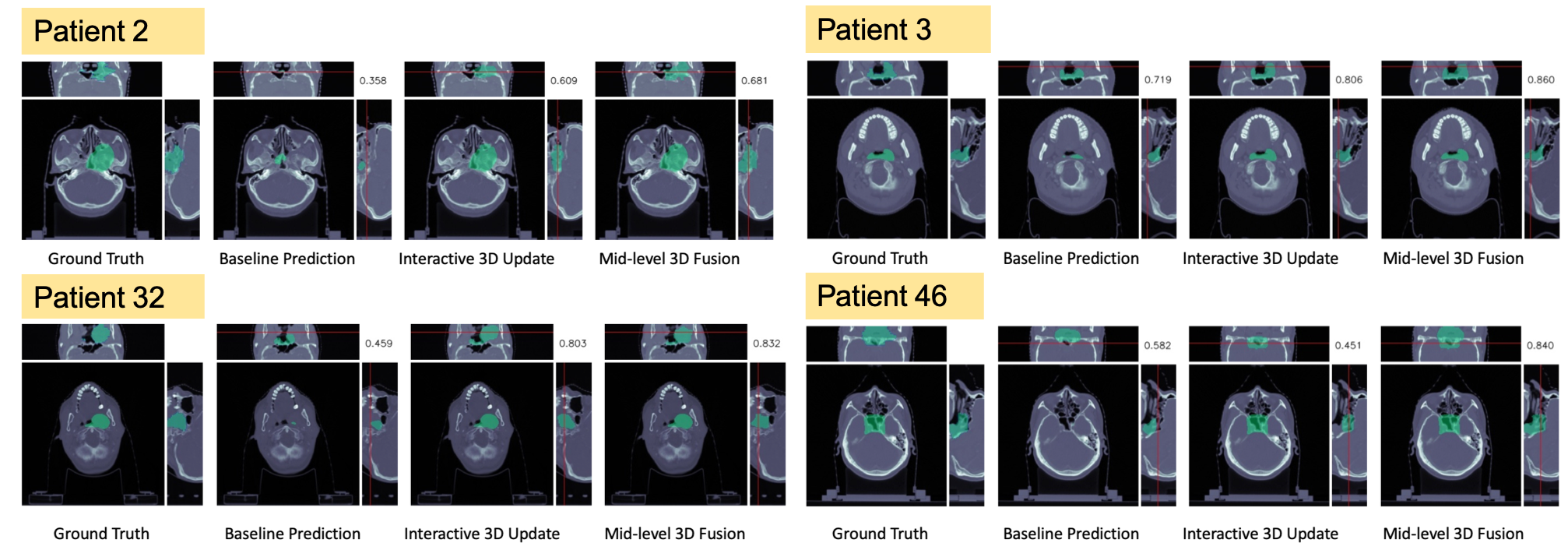}
\caption{Qaulitative result: Nasopharyngeal GTV. The figure demonstrate the GTV result of different methods in resized axial, sagittal, and coronal view. The red lines denote the location of the (worst) edited slice. The top right numbers denote the corresponding DSC scores.}
\label{fig:naso_qual}
\end{figure*}
\begin{figure*}[ht!]
\begin{center}
\includegraphics[width=0.99\linewidth]{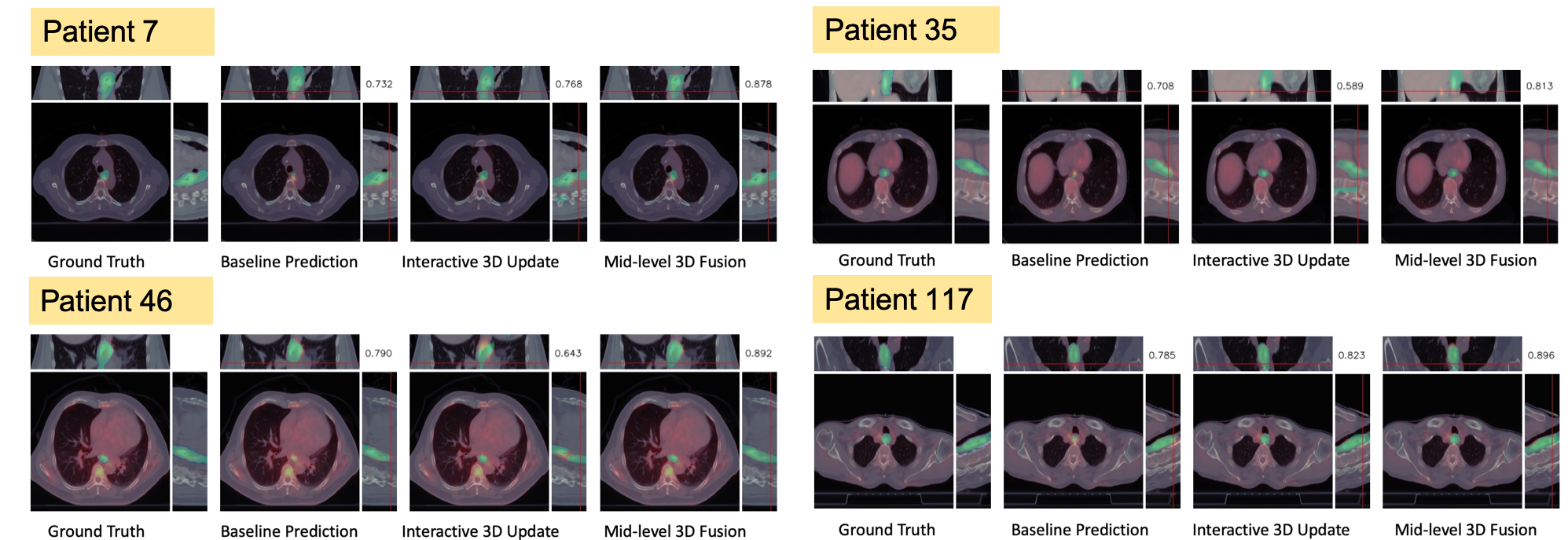}
\caption{Qaulitative result: Esophageal GTV. The figure demonstrate the GTV result of different methods in resized axial, sagittal, and coronal view. The red lines denote the location of the (worst) edited slice. The top right numbers denote the corresponding DSC scores.}
\label{fig:eso_qual}
\end{center}
\end{figure*}

\subsection{Qualitative Visualization}
To compare our result qualitatively with the baseline, we demonstrate several visualized single-slice-edited results on both Naso GTV and Eso GTV datasets in Fig.~\ref{fig:naso_qual} and Fig.~\ref{fig:eso_qual}. In both figures, the baseline prediction is the worst performed slice from the inference of original trained model $\phi_{seg}$. With the worst slice edited, our interactive 3D updated $\phi^{'}_{seg}$ is able to notably amend the sub-volume near the worst slice and increase the DSC performance. By using mid-level 3D fusion $\phi_{fuse}$, the prediction of far slices significantly improved while still preserving the nearby slices' prediction. Please refer to sagittal and coronal views for clear comparison. On the Eso GTV dataset shown in Fig.~\ref{fig:eso_qual} however, the prediction corruption cause by $\phi^{'}_{seg}$, as mentioned in Fig.~\ref{fig.bp_corruption}, become more severe than the Naso cases. Because of the larger receptive field needed for longer structure, the $\phi^{'}_{seg}$ prediction corrupted and get lower DSC. Fortunately, our $\phi_{fuse}$ is able to consider both nearby and far slices for edited slice propagation, generating better prediction than the baseline.

\subsection{Ablation Study}

\begin{figure*}[ht!]
\begin{center}
\includegraphics[width=0.75\linewidth]{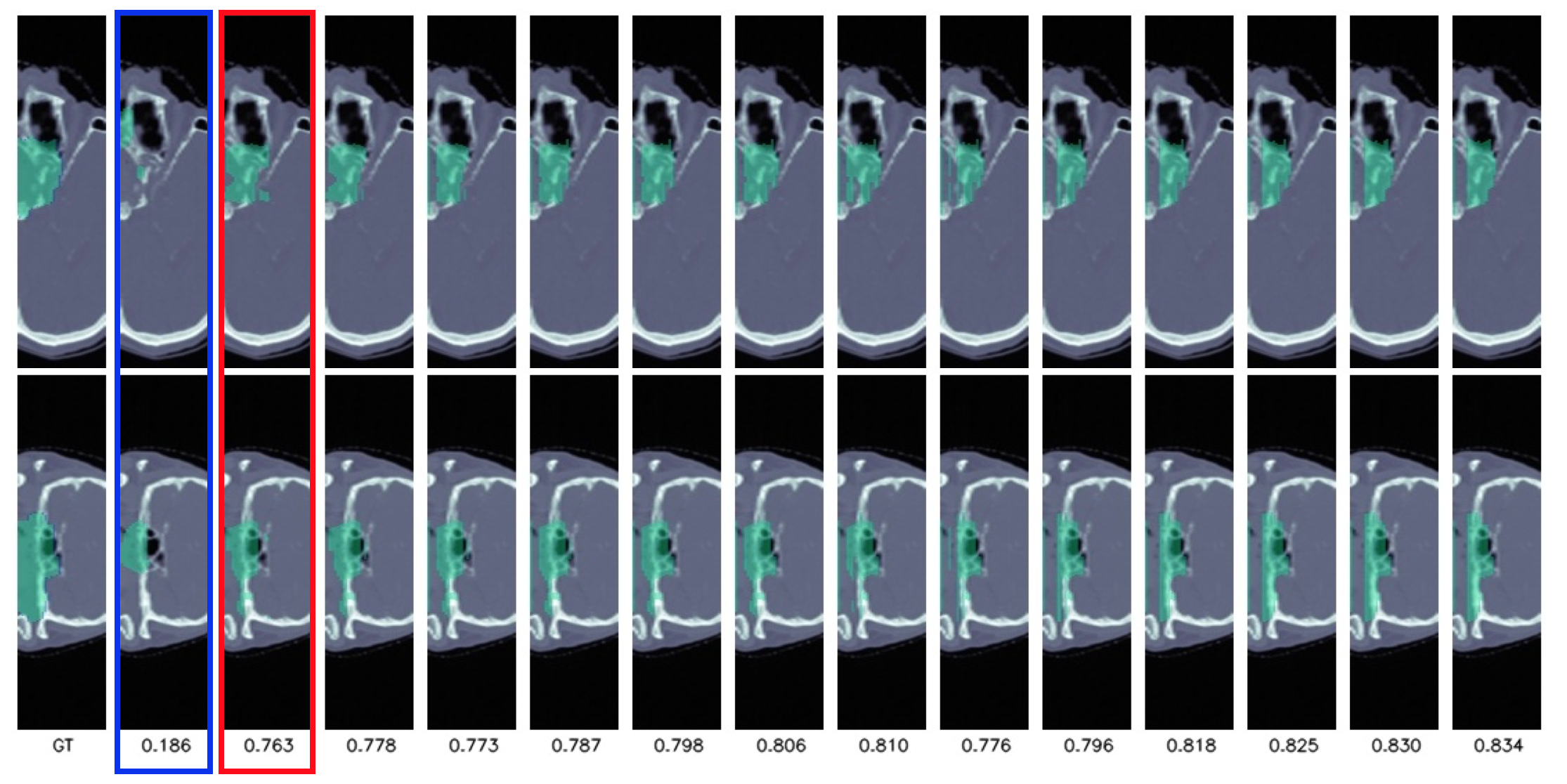}
\caption{Qaulitative result: Different steps of feature map update of a Nasopharyngeal GTV case. The figure demonstrate the GTV result of different steps from left to right in resized sagittal view (top) and coronal view (bottom) with corresponding DSC scores. Sequence order: GT, no update (blue), single update (red), etc.)}
\label{fig:ablation_naso_qual}
\end{center}
\end{figure*}

\begin{figure*}[!ht]
\begin{center}
\includegraphics[width=0.75\linewidth]{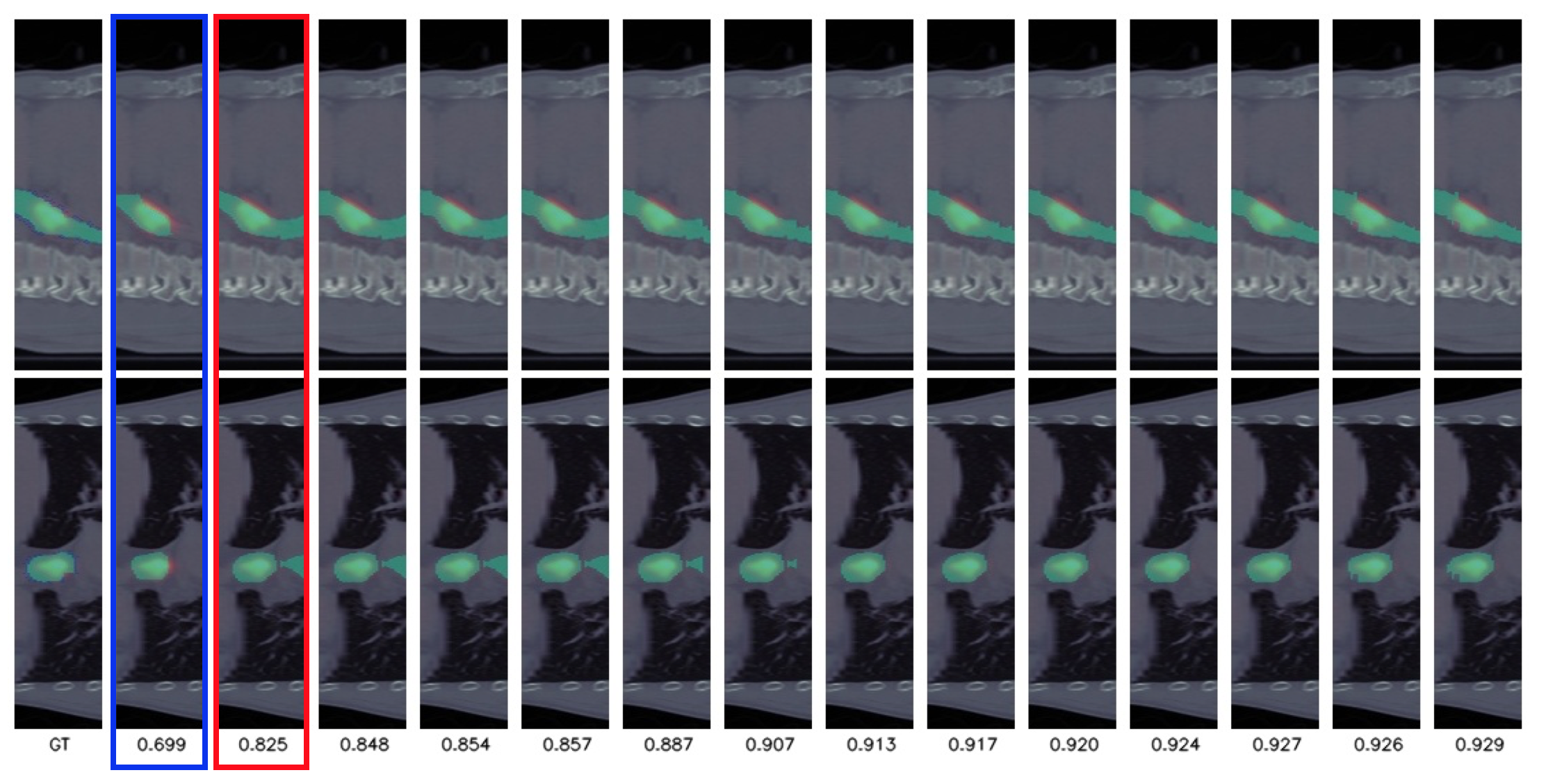}
\caption{Qaulitative result: Different steps of feature map update of a Esophageal GTV case. The figure demonstrate the GTV result of different steps from left to right in resized sagittal view (top) and coronal view (bottom) with corresponding DSC scores. Sequence order: GT, no update (blue), single update (red), etc.}
\label{fig:ablation_eso_qual}
\end{center}
\end{figure*}

In this section, we study the variation of performance under different settings of our method to demonstrate our design choices of each components.

\noindent\textbf{Various decoder level for interactive 3D update.}
Given one edited slice of the 3D prediction from features $\theta_{seg}$, our interactive 3D update module update it to $\theta_{seg}^{'}$ together with its prediction via back-propagation by activation. Choosing different level of $\theta_{seg}$ decoder will then cause different receptive field during the update. By choosing less layers, the updated receptive field will be small thus affecting less prediction slices; on the contrary, the more chosen layers, the larger the updated receptive field will be thus changing more prediction slices. Table.~\ref{tab.nasogtv-decoder} and Table.~\ref{tab.esogtv-decoder} indicate the DSC result of validation set on Naso GTV and Eso GTV dataset under at most 3 levels of chosen decoders. As shown in the tables, on both Naso GTV and Eso GTV dataset, updating 2 levels of decoder will have the best performance. As the result, we choose updating 2 levels of decoder as our final setting in order to change local slices' prediction near the edited slice while not affecting the far slices.

\begin{figure}[!h]
\begin{center}
\includegraphics[width=0.9\linewidth]{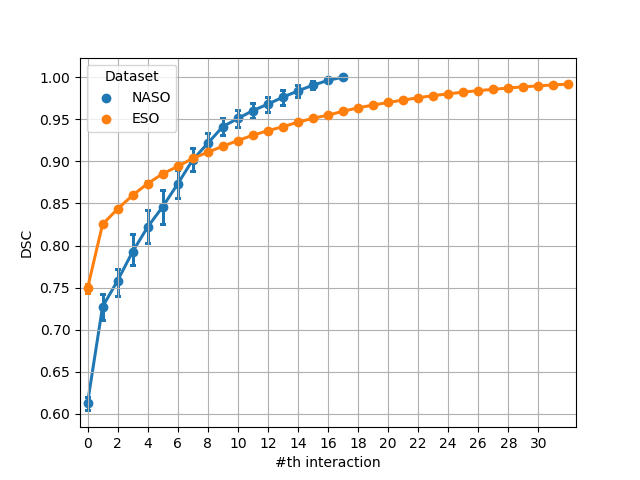}
\caption{Eso GTV DSC Score of Multiple interactive Steps.} 
\label{fig:multistep}
\end{center}
\end{figure}

\noindent\textbf{Multiple interactive steps.}
In addition to updating the model prediction by only one edited slice, our method is able to iterate the updating process multiple times. By updating multiple times by more edited slices, the prediction mask will become better but cost more time, annotation, and computation effort. Fig.~\ref{fig:multistep} illustrate the performance improvement with corresponding updating steps. As shown in the figures, the most significant improvement, about 10\% on both Naso GTV and Eso GTV dataset, occur when updating by the first edited slice (worst slice). The quality of prediction then gradually increases by more updating steps. Fig.~\ref{fig:ablation_naso_qual} and Fig.~\ref{fig:ablation_eso_qual} further illustrates the qualitative result under different number updating steps. Our method provides a flexible interactive way for different requirements during inference time for trade-off of performance and slice editing.



\begin{table}[h!] 
\caption{Nasopharyngeal GTV - Various Decoder Levels.}
  \centering
    \small
    \begin{tabular}{|c|p{1.8cm}|p{1.8cm}|p{1.8cm}|p{1.8cm}|}
    \hline
    Decoder Level & 1 & 2 & 3 \\
    \hline
    \hline
    DSC & 0.711$\pm$0.026 & 0.716$\pm$0.047 & 0.705$\pm$0.048  \\
    \hline
    \end{tabular}%
  \label{tab.nasogtv-decoder}%
\end{table}%

\begin{table}[h!] 
\caption{Esophageal GTV - Various Decoder Levels.}
  \centering
    \small
    \begin{tabular}{|c|p{1.8cm}|p{1.8cm}|p{1.8cm}|p{1.8cm}|}
    \hline
    Decoder Level & 1 & 2 & 3 \\
    \hline
    \hline
    DSC & 0.865$\pm$0.017 & 0.872$\pm$0.005 & 0.870$\pm$0.003  \\
    \hline
    \end{tabular}%
  \label{tab.esogtv-decoder}%
\end{table}%



\section{Conclusion}
In this paper, we present an interactive 3D radiotherapy target refinement framework is proposed. Our proposed method leverages the backpropagation for activation technique to transport the user input backwardly into the latent space which intrinsically influence larger scope along z-axis and a fusion network then integrates the original and the updated high level features to generate new applicable features for the following decoder to produce revised prediction volume according to the user-edited image slice. Our interaction framework only requires radiation oncologists to edit a small amount of poorly predicted slice in the whole prediction volume to yield the satisfactory GTV volume and is compatible with most main stream 3D segmentation networks without retraining or modifying their architectures. Extensive experiments on two published datasets, the Nasopharyngeal Cancer GTV dataset and the Esophageal Cancer GTV dataset, and their corresponding segmentation architectures are conducted to demonstrate our method is applicable to disparate cancers, image modalities, and backbone networks. The results show that the proposed method is capable of are able to further improve the state-of-the-art prediction results with sparse user interactions for treatment planning for different cancers.
\label{sec:conclusion}

\bibliographystyle{IEEEtran}
\bibliography{bibliography}

\end{document}